\documentclass[12pt,preprint,notoc,nohyper]{MCsDefntn} 

\usepackage{epsfig,multicol,bbm}

\newcommand\fverb{\setbox\pippobox=\hbox\bgroup\verb}
\newcommand\fverbdo{\egroup\medskip\noindent%
            \fbox{\unhbox\pippobox}\ }
\newcommand\fverbit{\egroup\item[\fbox{\unhbox\pippobox}]}
\newbox\pippobox

\title{On the definition of matter collineations}

\author{Asghar Qadir$^{a}$ and K. Saifullah$^b$ \\

$^a$Centre for Advanced Mathematics and Physics \\ National
University of Sciences and Technology, Rawalpindi, Pakistan \\
$^b$Department of Mathematics, Quaid-i-Azam University, Islamabad,
Pakistan (Electronic address: aqadirmath@yahoo.com,
saifullah@qau.edu.pk) \\}

\preprint{}  

\abstract{It is shown that when the stress-energy tensor of a
spacetime is diagonal and is written in the mixed form, its
collineations admit infinite dimensional Lie algebras except
possibly in the case when the tensor depends on all the spacetime
coordinates. The result can be extended for more general second rank
tensors.}

\dedicated{}

\begin{document}

Symmetries of the spacetime metric, ${\bf g}$, are given by the
Killing equations
\begin{equation} \pounds_{\bf \xi} {\bf g} = 0, \label{isom}
\end{equation}
where $\pounds_{\bf \xi}$ represents the Lie derivative and
$\bf{\underline{\xi}}$ are called {\it Killing vectors} (KVs). If
${\bf g}$ in the above equation is replaced by some other tensor the
symmetry vectors are called {\it collineations} of that
tensor\cite{katzin,hallbook,duggal,2r}. Thus, for example, we define
Ricci collineations (RCs) by

\begin{equation}
\pounds _{\mathbf{\xi}}\mathbf{R}=0\; . \label{f17}
\end{equation}

When we write the Ricci tensor, ${\bf R}$, in covariant form, the
above equation can be written in components as

\begin{equation}
\xi^{c}R_{ab,c}+R_{ac} \xi_{,b}^{c}+R_{bc} \xi_{,a}^{c}=0\, .
\label{f18}
\end{equation}

If ${\bf R}$ in the above equation is replaced by ${\bf T}$, the
stress-energy tensor, $\bf{\underline{\xi}}$ is called a matter
collineation (MC) \cite{carot,hall}. Similarly we define Weyl and
curvature collineations. Applying these symmetry constraints is one
way to obtain solutions of the system of Einstein field equations
(EFEs)

\begin{equation}
R_{\mu\nu} - \frac{1}{2} R g_{\mu\nu} = \kappa T_{\mu\nu} ,
\end{equation}
which is highly non-linear and is very difficult to solve otherwise.
Ricci colli- neations (RCs), for example, have been used to solve
EFEs\cite{QSZ}. These symmetries lead to conservation laws also.
Davis \emph{et al.}\cite{DK1,DK2} did the pioneering work on the
important role of RCs and the related conservation laws that are
admitted by particular types of matter fields. They showed that the
existence of isometries and collineations leads to conservation laws
in the form of integrals of a dynamical system. They also considered
the application of these results to relativistic hydrodynamics and
plasma physics. Oliver and Davis\cite{OD} obtained conservation
expressions for perfect fluids using RCs. The properties of fluid
spacetimes admitting RCs were studied by Tsamparlis and
Mason\cite{TM}. They have studied perfect fluid spacetimes in detail
and have also considered a variety of imperfect fluids with
cosmological constant and with anisotropic pressure. Apart from
these physical considerations symmetries provide an invariant basis
for classification of spacetimes also (see, for example, Refs.
\cite{QSZ}, \cite{BQZ}-\cite{cb2}).

To actually compute the collineations one needs to write the
equations for them in component form (\cite{QSZ},
\cite{BQZ}-\cite{cb2}). Thus the valence\cite{penroseRR} of the
tensor is relevant. In fact, the symmetry algebra is different for
the different forms, in general. While it is ``natural" to use the
covariant forms for the metric and the Ricci tensor, ${\bf R}$, for
the stress-energy tensor, $\bf{T}$, it is far from obvious which
form should be used. However, MCs generally discussed in the
literature\cite{KS1,KS2,cb1,cb2} are obtained from the covariant
form. The question as to whether one should use the contravariant
($T^{\mu\nu}$), covariant ($T_{\mu\nu}$) or mixed ($T^{\mu}_{\nu}$)
form of the stress-energy tensor, becomes more important when one
notes that all these forms have their own separate significance in
general relativity. While the original field theoretic definition of
the tensor comes from\cite{landau}

\begin{equation}
T^{\mu}_{\nu}=q^{\alpha}_{,\nu}\frac{\delta L}{\delta
q^{\alpha}_{,\mu}}- \delta^{\mu}_{\nu}L ,
\end{equation}
where $L$ is the Lagrangian density and $q^{\alpha}$ are the
generalized coordinates, it is the covariant form that arises
naturally in the EFEs. On the other hand, the contravariant form is
needed for the Hamiltonian formulation\cite{MTW}. This ambiguity in
the definition of MCs has also been pointed out in Ref.~\cite{hall}.

There are two questions to be considered: (a) which (if any) of the
forms can be regarded as ``correct"; (b) what can be obtained from
each of them? We consider the form obtained from the field theoretic
definition and explore question (b). In particular, we show that for
a diagonal mixed form stress-energy tensor the Lie algebra for the
MC vectors will be infinite dimensional except when it depends on
all four spacetime variables in which case it can be finite
dimensional. When we say that $T^{\mu}_{\nu}$ is diagonal we mean
that it is of the diagonal Segre type\cite{2r} and it is diagonal
when expressed as a matrix in coordinate components. Further, the
MCs are assumed to be smooth, so that they form a Lie algebra. The
MC equations in component form are

\begin{equation} \label{mc}
\xi^\rho T^{\mu}_{\nu,\rho} + T^{\mu}_{\rho} \xi^{\rho}_{,\nu} -
T^{\rho}_{\nu} \xi^\mu_{,\rho} = 0 .
\end{equation}
Here the components of $T^{\mu}_{\nu}$ and the MC vector,
$\xi^{\mu}$,  are functions of the coordinates $x^{\mu}$. We note
that while in the covariant (or contravariant)
form\cite{KS1,KS2,cb1,cb2} there are ten equations in four
dimensions, in the mixed form we have sixteen equations. It is worth
mentioning here that if $T^{\mu}_{\nu}$ is \emph{a} tensor (as will
be discussed later) then these sixteen equations are independent,
but if it represents the stress-energy tensor then they are
restricted by $T^{\mu}_{\nu} g_{\mu \rho}=T^{\mu}_{\rho}
g_{\mu\nu}$. Dropping the summation convention here and hereafter,
for the diagonal stress-energy tensor equation (\ref {mc}) can be
written as

\begin{equation} \label{12eqn}
\xi^{\mu}_{,\nu}(T^{\mu}_{\mu} - T^{\nu}_{\nu}) = 0 , \, \qquad
(\mu, \nu = 0,1,2,3) \,
\end{equation}
which is a set of 12 equations, for $\mu \neq \nu$, and
\begin{equation} \label{4eqn}
\sum_{\rho=0}^{3} \xi^{\rho}T^{\mu}_{\mu,\rho}= 0
\end{equation}
is a set of 4 equations, as it applies for each $\mu$ separately.

Note that in the first set of twelve equations if $T^{\mu}_{\mu}\neq
T^{\nu}_{\nu}$ ($\mu\neq\nu$) then

\begin{equation} \label{1}
\xi^{\mu}_{,\nu} = 0 ,
\end{equation}
that is, if all the components of the stress-energy tensor are
different then for each $\mu$, $\xi^\mu$ can only be a function of
$x^{\mu}$. For the next set of four equations (\ref{4eqn}) we do the
analysis by viewing $T^{\mu}_{\mu , \nu}$ as the matrix $\mathbf{A}$
with elements $A^{\mu}_{\nu}$. Now, if $rank(\mathbf{A})=4$, then
$\xi^\mu=0$, i.e. the columns of $\mathbf{A}$ are 4 linearly
independent vectors in a 4-dimensional space and
$\mathbf{\underline{\xi}}$ is orthogonal to all of them, and hence,
it is a zero vector. Now, if $rank(\mathbf{A})$ is less than 4, then
$\mathbf{\underline{\xi}}$ is an eigenvector of $\mathbf{A}$ with
zero eigenvalue and the subspace spanned by these eigenvectors
satisfy equations (\ref{4eqn}). Thus, for example, if
$rank(\mathbf{A})$ is 3, then there will be $4-3=1$ such
eigenvectors and we can say that the vector has one degree of
freedom, or the MC algebra can be written\cite{penroseRR} as
$\infty^{1}$. Note that if $\xi^a$ is a solution of equations
(\ref{4eqn}) then for some function $f$, $f\xi^a$ is also a solution
but this will be subject to equations (\ref{12eqn}) which may only
increase the degree of freedom. Further, if $T^{\mu}_{\nu}$ depends
on one variable only then the rank of $\mathbf{A}$ is 1 at most and
degrees of freedom will at least be $\infty^{3}$. If it depends on
two variables the maximum rank will be 2 and degrees of freedom will
at least be $\infty^{2}$; if three variables the maximum rank is 3.
Similarly, if the rank is 0 (for example, when the tensor is
constant) it is $\infty^{4}$.

In the above we assumed that all the components of $T^{\mu}_{\mu}$
are different. If we drop this assumption we see that, when
equations (\ref{12eqn}) are imposed on the above results different
possibilities arise, but in all these cases the degree of freedom
will only increase.

In order to show that a finite dimensional algebra is indeed
admitted, we construct an example. For non-trivial collineation
vector with finite dimensional Lie algebras to exist it is necessary
that the rank of the matrix ($T^{\mu}_{\mu,\nu}$) be 3. For example,
take

$$T^{0}_{0} = t+x+y , \, \qquad T^{1}_{1}=t+x+z ,$$
$$ T^{2}_{2} = t+y+z , \, \qquad T^{3}_{3} =x+y-2z ,$$
where $x^{0}=t$, $x^{1}=x$, $x^{2}=y$, $x^{3}=z$ are arbitrary
coordinates and not necessarily Cartesian coordinates. The solution
is

\begin{equation}
\mathbf{\xi }= -2\frac{\partial}{\partial
t}+\frac{\partial}{\partial x}+\frac{\partial}{\partial
y}+\frac{\partial}{\partial z} \,,
\end{equation}
which is non-trivial. Note that for simplicity we have assumed units
in which the tensor can be given in the same units as the
coordinates. More generally, appropriate constants could have been
inserted.

Thus we conclude that \emph{for a diagonal stress-energy tensor
written in the mixed form, the MCs (defined by equations (\ref{mc}))
admit an infinite dimensional Lie algebra except possibly when the
tensor depends on all the spacetime coordinates. If
$rank(\mathbf{A})$ is zero, the minimum degrees of freedom are 4; if
one, the minimum is 3; two, 2; three, 1. If $rank(\mathbf{A})=4$ the
MC vector is zero.}

Let us assume that for a class of spacetimes there exists a
coordinate system in which one or more of the coordinates are
missing from all the components of the tensor. Then the algebra will
become infinite dimensional. Thus in the case of static spacetimes,
for example, the Lie algebras for diagonal mixed MC vectors are
infinite dimensional. For example  Refs.~\cite{KS1,KS2,cb1,cb2}
dealing with static spacetimes use the covariant form, $T_{\mu
\nu}$, and obtain some finite dimensional Lie algebras. Here we see
that they are all infinite dimensional Lie algebra for the mixed
form $T^{\mu}_{\nu}$.

It is worth mentioning here that while we considered the
stress-energy tensor, the results hold for any tensor of rank two in
a space of any dimensions. Thus we can state the following result.

\emph{For a diagonal tensor of rank two written in the mixed form,
the Lie symmetries (as defined by equations (\ref{mc})) have
infinite dimensional Lie algebras except possibly when the tensor
depends on all the coordinate variables. The maximum number of
degrees of freedom is $n^2$ and the minimum zero, where $n$ is the
dimension of the space.}

For example, consider the Friedmann-Robertson-Walker spacetime in
stereographic coordinates (for $k=0$, i.e. the flat case)
\begin{equation}
ds^2 = d\tilde{t}^2 - S(\tilde{t})^2 (dx^2 +dy^2+ dz^2) ,
\end{equation}
where $S$ is a non-zero function of $\tilde{t}$. Its stress-energy
tensor is
\begin{eqnarray}
\left.
\begin{array}{l}
T_{0}^{0}=3\dot{S}^2/S^2=A(\tilde{t}) , \\
T_{1}^{1}=T_{2}^{2}=T_{3}^{3}=-\left( \dot{S}^2 + 2 S
\ddot{S}\right)=B(\tilde{t}) .
\end{array}
\right.
\end{eqnarray}

The MCs for this spacetime for the covariant (or contravariant)
stress-energy tensor form the following ten dimensional Lie
algebra\cite{cb1}

\begin{eqnarray}
\left.
\begin{array}{l}
\mathbf{X}_1 = \partial_x, \\
\mathbf{X}_2 = \partial_y, \\
\mathbf{X}_3 = \partial_z, \\
\mathbf{X}_4 = z \partial_x - x \partial_z, \\
\mathbf{X}_5 = y \partial_z -z \partial_y, \\
\mathbf{X}_6 = x \partial_y - y \partial_x, \\
\mathbf{X}_7 = \partial_t, \\
\mathbf{X}_8 = t \partial_x - \epsilon x \partial_t, \\
\mathbf{X}_9 = t \partial_y - \epsilon y \partial_t, \\
\mathbf{X}_{10} = t \partial_z - \epsilon z \partial_t ,
\end{array}
\right.
\end{eqnarray}
where $\mathbf{X}_7, \ldots, \mathbf{X}_{10}$ are
\emph{non-isometric} MCs. In the above $\epsilon =+1$ or $-1$, and
$dt =\sqrt{A/B} d\tilde{t}$. Further, we have assumed that $B$ is a
constant, but even if we drop this assumption, we again get a ten
dimensional algebra\cite{cb1}. In contrast to this, if we use the
mixed form of the stress-energy tensor, equations (\ref{mc}) do not
constrain the vector sufficiently and we get the following form of
the MC vector with infinite degrees of freedom

\begin{eqnarray}
\left.
\begin{array}{l}
\xi^{0}= 0\,,  \\
\xi^{1}=f(x, y, z) \,, \\
\xi^{2}=g(x, y, z) \,, \\
\xi^{3}=h(x, y, z) \,, \\
\end{array}
\right.
\end{eqnarray}
where $d$, $f$, $g$ and $h$ are arbitrary functions. However, if the
$T^{\mu}_{\nu}$ are independent of the time coordinate, we get
$\xi^{0}=d(t)$, $S$ becomes constant and spacetime turns out to be
Minkowski (and we again have an infinite dimensional algebra).

Finally we remark that when the stress-energy tensor is written in
the covariant form its symmetries can be compared with those of the
Ricci tensor (the RCs)\cite{KS1,KS2}. (In this case it may be more
appropriate to call them \emph{Einstein collineations} as they are
essentially collineations of the Einstein tensor, by the Einstein
Field Equations.) For the contravariant form the number of
independent equations being the same as in the covariant form, the
number of symmetry generators would be the same. It is worth
checking whether the collineations are directly related (through
constants or at most the metric coefficients as linear combinations)
or not. In principle the presence of the metric tensor could
introduce major differences.

Further, the dimensionality of the Lie algebra of collineations also
depends on the degeneracy of a tensor. The metric tensor is
non-degenerate, therefore, its symmetries are always finite
dimensional. When the stress-energy tensor is written in covariant
form it always has a finite dimensional Lie algebra if it is
non-degenerate (i.e. when $det{T_{ab}}\neq0$). As such, the results
on MCs of the non-degenerate tensor in Ref.~\cite{sharifer} (cases
on pages 5151, 5152 and four cases in Table III) are obviously
incorrect. Similarly, the infinite dimensional Lie algebras for the
non-degenerate Ricci tensor claimed in Ref.~\cite{ziader} (Theorems
8, 9, 11 and 12, and the corresponding claim in the Abstract) also
cannot be correct.

Notice that different valences of the stress-energy tensor
correspond to different symmetry algebras. One needs to select the
definition of matter collineations according to the physical
application. For example, to construct conservation laws in the
Hamiltonian formalism, symmetries of the contravariant tensor would
be most appropriate.

It is worth mentioning that there is also an ambiguity in the
definition of collineations for tensors of rank four. In particular,
Weyl collineations\cite{ibrar} require more detailed analysis as the
Weyl tensor could be defined as the trace-free part of the curvature
tensor\cite{hallbook}, in which case its valence would be
$\left(^{1}_{3}\right)$. Alternatively, as defined from spinor
considerations\cite {PenRin} it should be regarded as covariant,
i.e. of valence $\left(^{0}_{4}\right)$. For 4 dimensions, the
latter set is of 20 equations while the former is of $96-16-10=70$
equations (from the skew symmetry of the last pair of indices, the
first Bianchi identities and the fact that the Weyl tensor is
trace-free).

\acknowledgments The authors are grateful to Ugur Camci for his help
during the write-up of this paper. They also thank the 30th
International Nathiagali Summer College on Physics and Contemporary
Needs where this work was partly done. KS acknowledges a research
grant from the Higher Education Commission of Pakistan.

\end{document}